\documentclass[12pt]{article}
\usepackage{graphicx}

\def\Title#1{\begin{center} {\Large #1 } \end{center}}
\def\Author#1{\begin{center}{ \sc #1} \end{center}}
\def\Address#1{\begin{center}{ \it #1} \end{center}}

\newenvironment{Abstract}{\begin{quotation}  }{\end{quotation}}
\newenvironment{Presented}{\begin{quotation} \begin{center} 
             PRESENTED AT\end{center}\bigskip 
      \begin{center}\begin{large}}{\end{large}\end{center} \end{quotation}}
\begin{document}
\begin{titlepage}
 %\pubblock
\vfill
\Title{Overview of ATLAS Forward Proton (AFP) detectors in Run-2 and outlook for Run-3 analyses}
\vfill
\Author{Andr\'e  Sopczak, \\
on behalf of the ATLAS Forward Detectors group}
\Address{Institute of Experimental and Applied Physics, \\ Czech Technical University in Prague}
\vfill
\begin{Abstract}
We describe the status of the ATLAS Forward Proton (AFP) detectors in Run-2 and the outlook for Run-3 analyses.
The performance is discussed. This includes the Tracking and Time-of-Flight detectors, 
the luminosity,
the alignment,
the trigger, 
and data quality monitoring.
Additionally, key physics results from the first AFP analyses are showcased.
\end{Abstract}
\vfill
\begin{Presented}
DIS2023: XXX International Workshop on Deep-Inelastic Scattering and
Related Subjects, \\
Michigan State University, USA, 27-31 March 2023 \\
     \includegraphics[width=9cm]{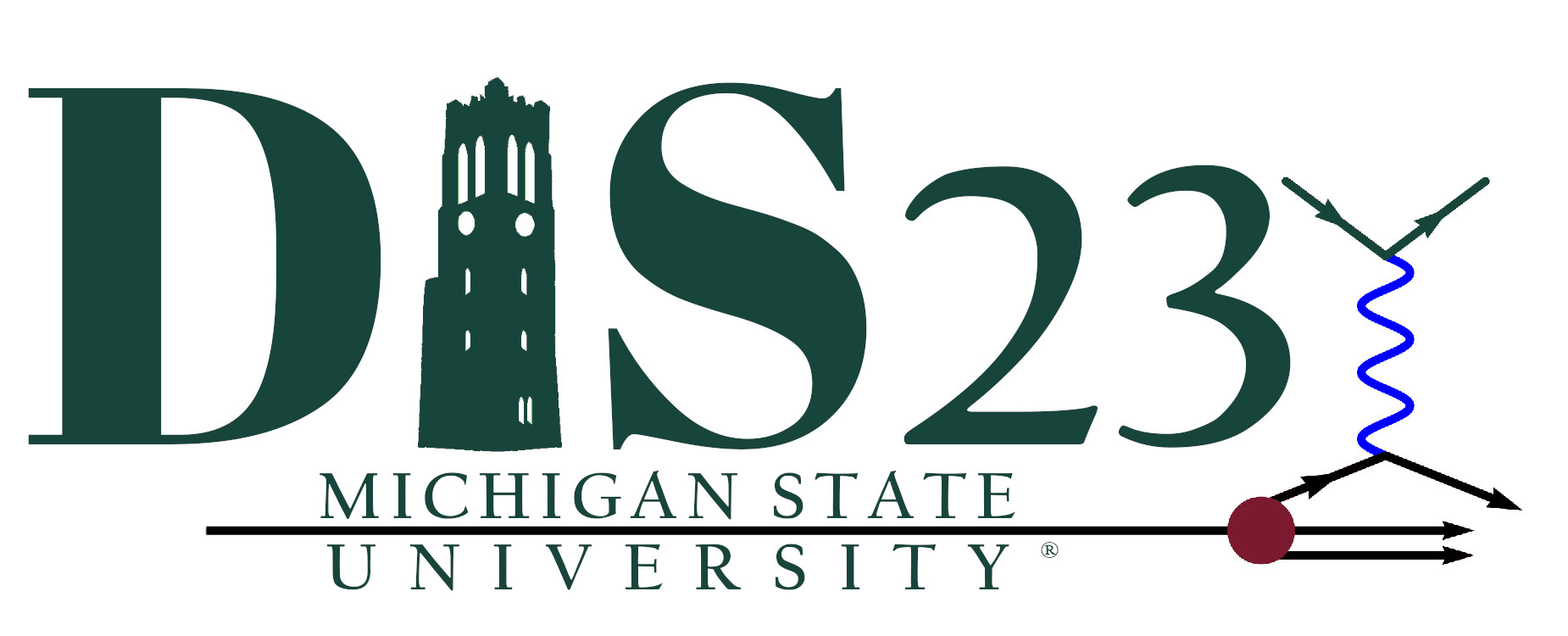}
\end{Presented}
\vfill
\end{titlepage}

\section{Status of ATLAS Forward Proton\,(AFP)\,detectors}
The AFP detectors are located at about $-200$\,m and $+200$\,m from the Interaction Point (IP) in the central ATLAS detector~\cite{AFPpage}.
For example in $\gamma\gamma \rightarrow \ell\ell$ or $\gamma\gamma \rightarrow\gamma\gamma$ events, the
final state proton can be intact and 
be record in the AFP detectors.
The AFP detectors took data during LHC Run-2 operation and they are taking data in Run-3.

%Each side of the AFP systems is referred to as an arm.
The Silicon Tracker (SiT) consists of four layers of silicon pixel detectors, two NEAR and two FAR stations.
Only FAR stations are equipped with Time-of-Flight (ToF) detectors~\cite{AFPpage}.
% \subsection{Silicon Tracker (SiT)}
% \subsection{Time of Flight (ToF)}

\vspace*{-0.2cm}
\section{Luminosity}
\begin{figure}[htb]
\vspace*{-0.3cm}
\begin{center}
\includegraphics[width=0.32\textwidth,trim=0cm 7.5cm 0.5cm 12cm]
{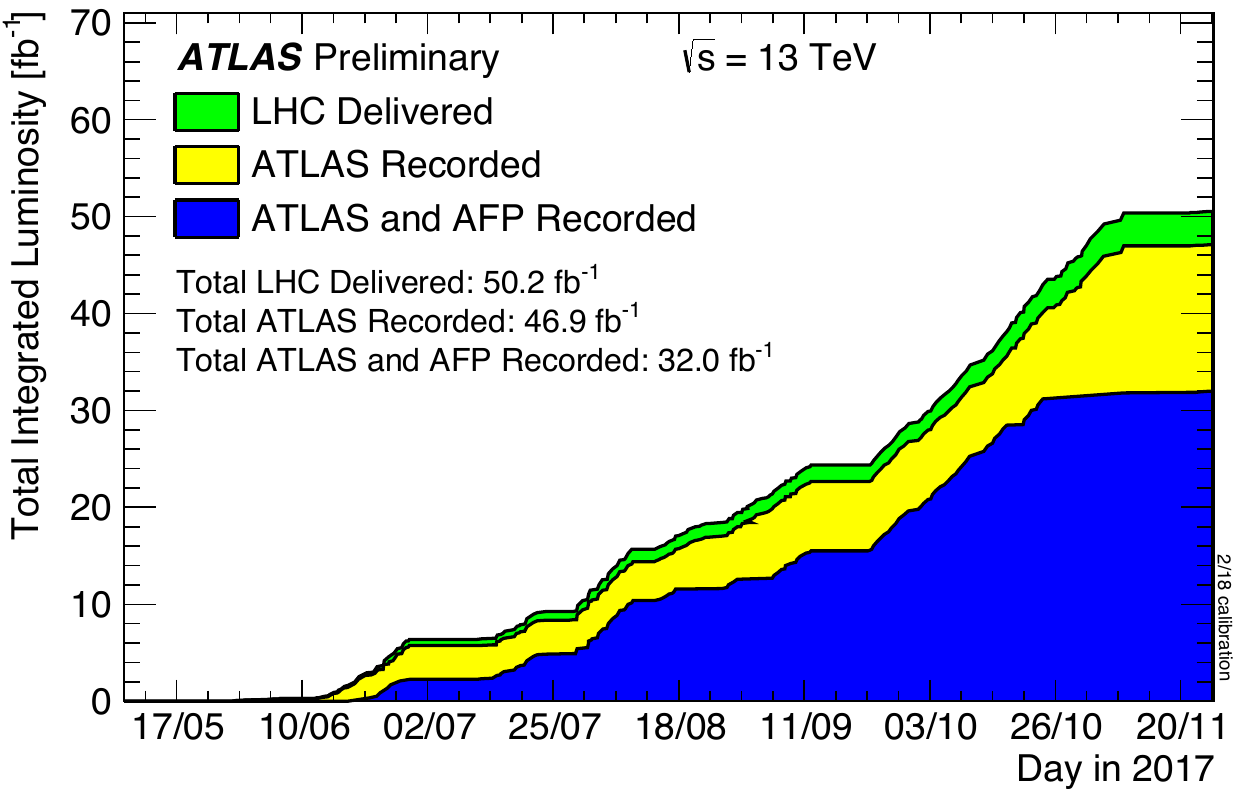}\hfill
\includegraphics[width=0.32\textwidth]
{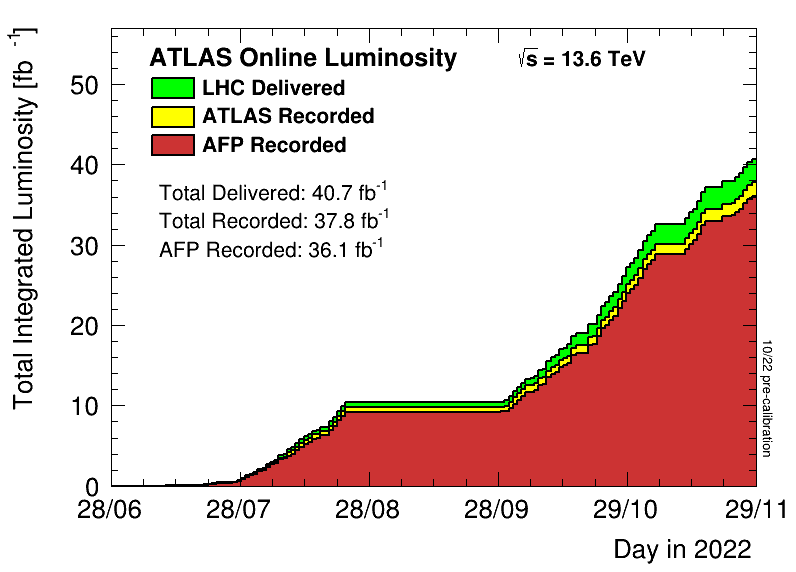}
\includegraphics[width=0.34\textwidth]
{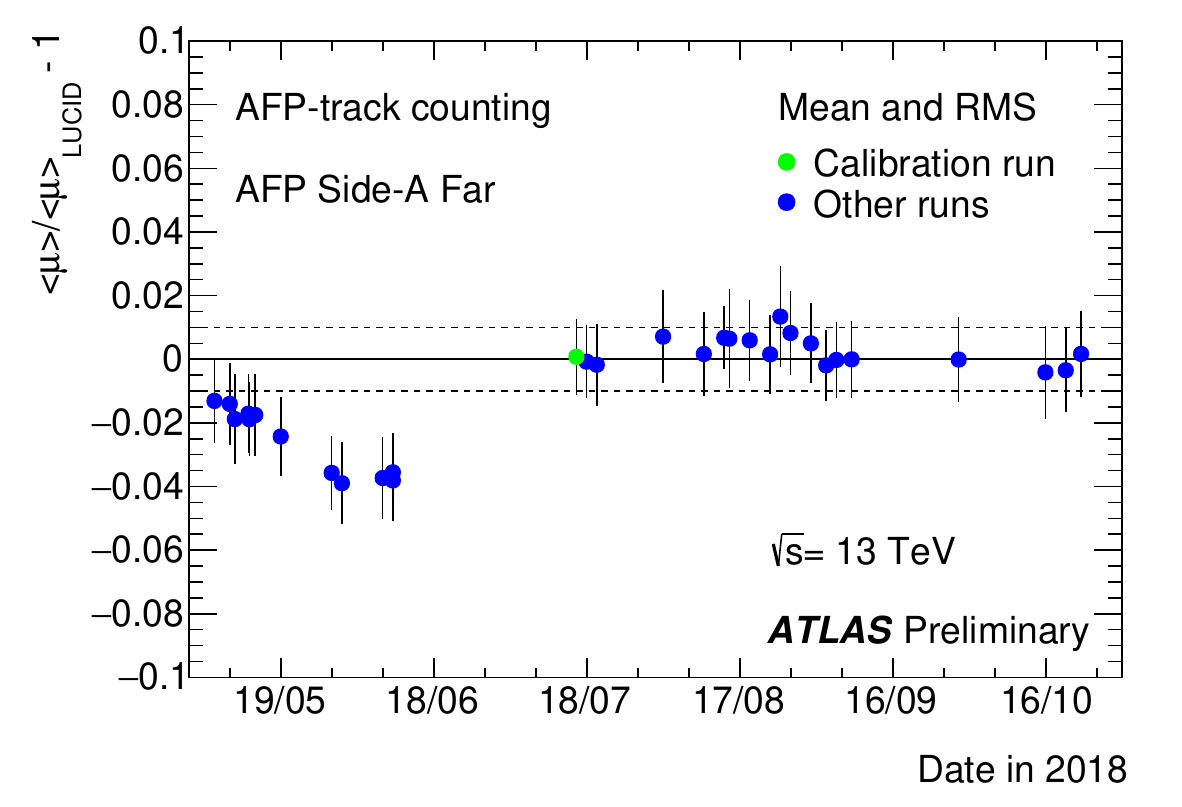}
\end{center}
\vspace*{-0.8cm}
\caption{
Luminosity 2017 (left), 2022 (center) and long-term AFP luminosity versus ATLAS reference luminosity, determined by the LUCID luminometer (right)~\cite{AFPpage,LumiRun3}.
}
\label{fig:lumi}
\vspace*{-0.7cm}
\end{figure}

%The AFP operated in 2017 and from 2022.
In 2017, 32.0\,fb$^{-1}$ data at 13\,TeV, and in 2022, 36.1\,fb$^{-1}$ data at 13.6\,TeV were taken in co-incidence with the ATLAS central detector (Fig.~\ref{fig:lumi}~\cite{AFPpage,LumiRun3}).

% https://twiki.cern.ch/twiki/bin/view/AtlasPublic/LuminosityPublicResultsRun3
% https://twiki.cern.ch/twiki/pub/AtlasPublic/ForwardDetPublicResults

In 2018 AFP data, the average number of interactions per bunch crossing, $\mu$, for each luminosity block (60\,s) was determined from the average number of AFP tracks.
From 31 runs of data-taking (12-24 hours), $\mu$ values were measured and compared to measurements by the ATLAS luminometer LUCID.
The agreement is within 1\% between July and October 2018
(Fig.~\ref{fig:lumi}~\cite{AFPpage,LumiRun3}).
% https://twiki.cern.ch/twiki/view/AtlasPublic/ForwardDetPublicResults

\section{SiT data-taking, hit map July 2022 LHC Run-3}
The SiT planes of the AFP detector are $336\times80$ (row $\times$ column) pixels with dimensions of $50\,\mu{\rm m} \times  250\,\mu{\rm m}$.
The hit map shows a clear correlation between SiT Raw ID and SiT Column ID, indicating the beam position (Fig.~\ref{fig:hitmap}~\cite{AFPpage}).
%https://twiki.cern.ch/twiki/view/AtlasPublic/ForwardDetPublicResults

\vspace*{-0.2cm}
\section{Alignment}
\begin{figure}[htb]
\vspace*{-0.3cm}
\begin{center}
\includegraphics[width=0.3\textwidth]
{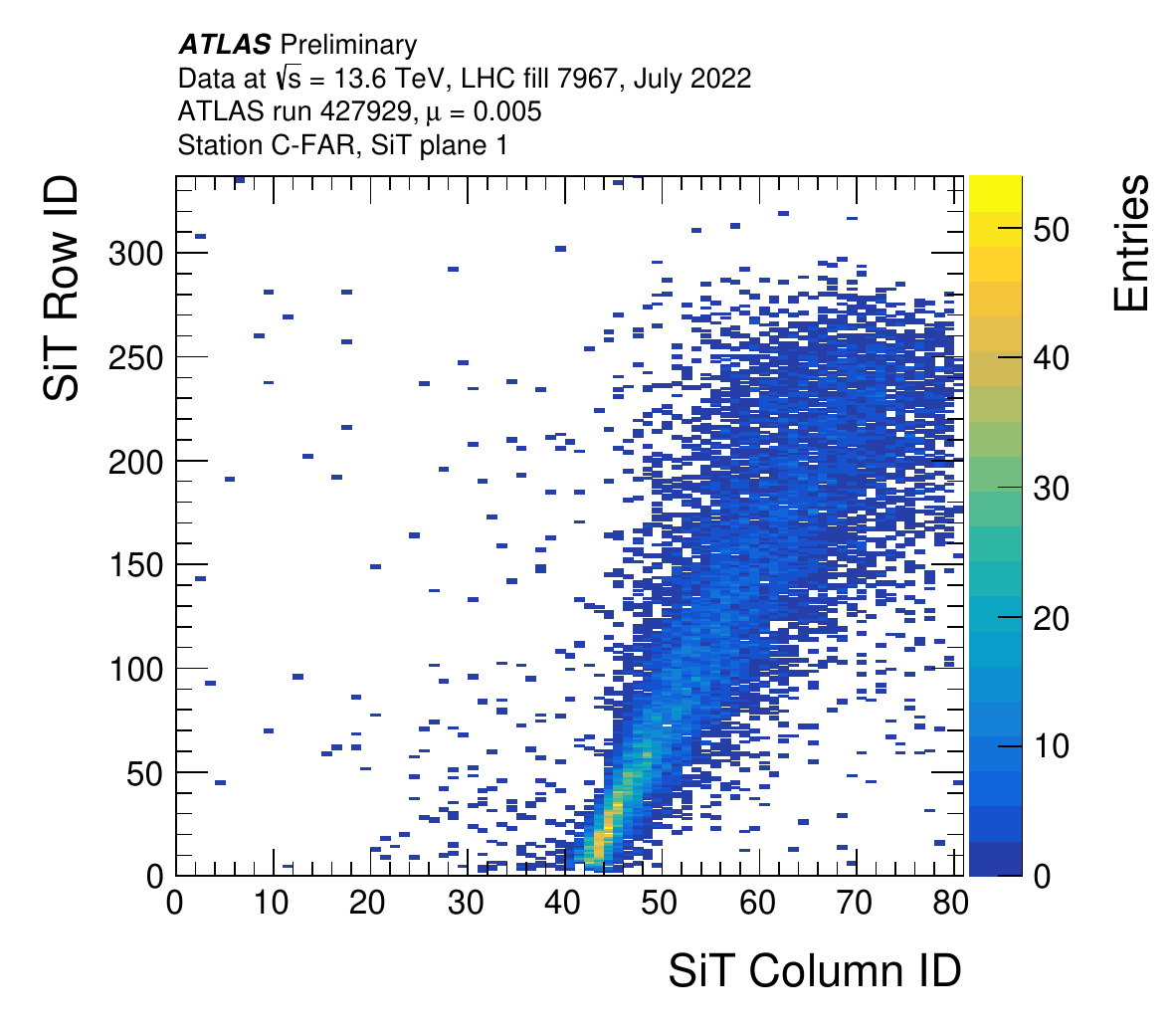}\hfill
\includegraphics[width=0.345\textwidth]
{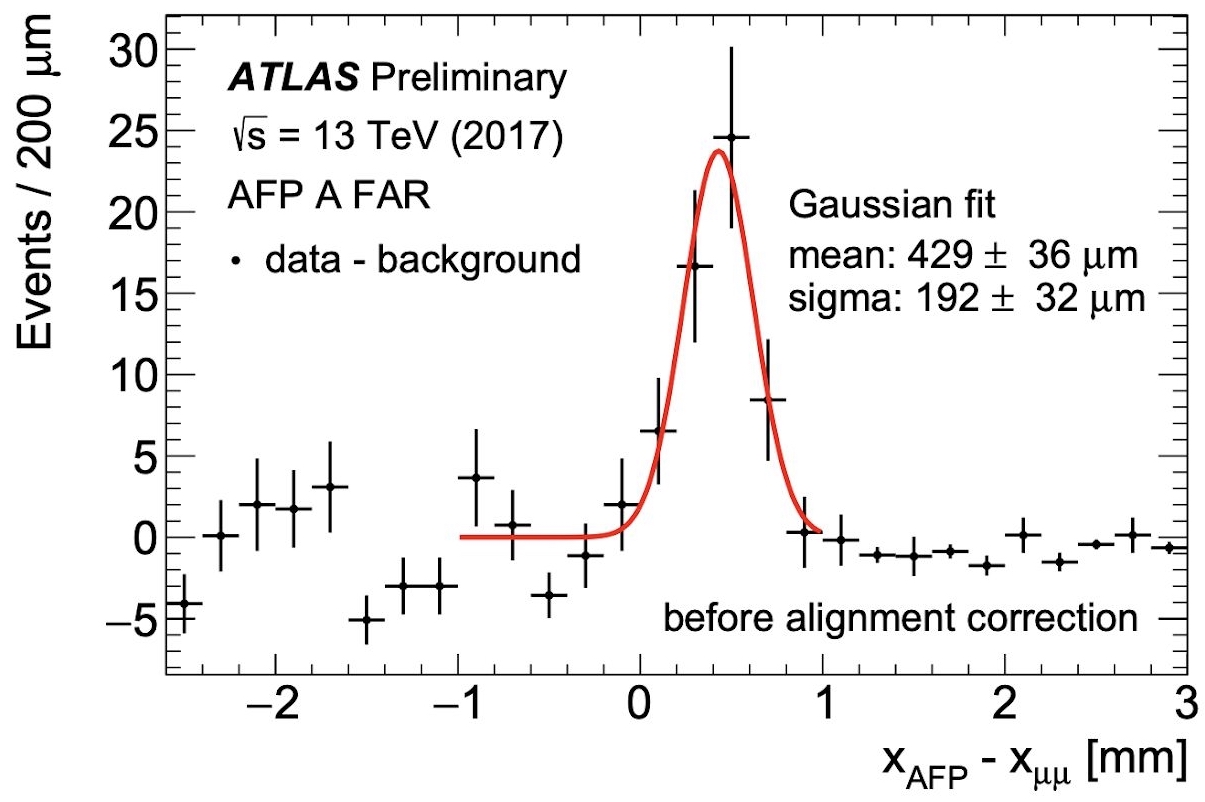}
\includegraphics[width=0.345\textwidth]
{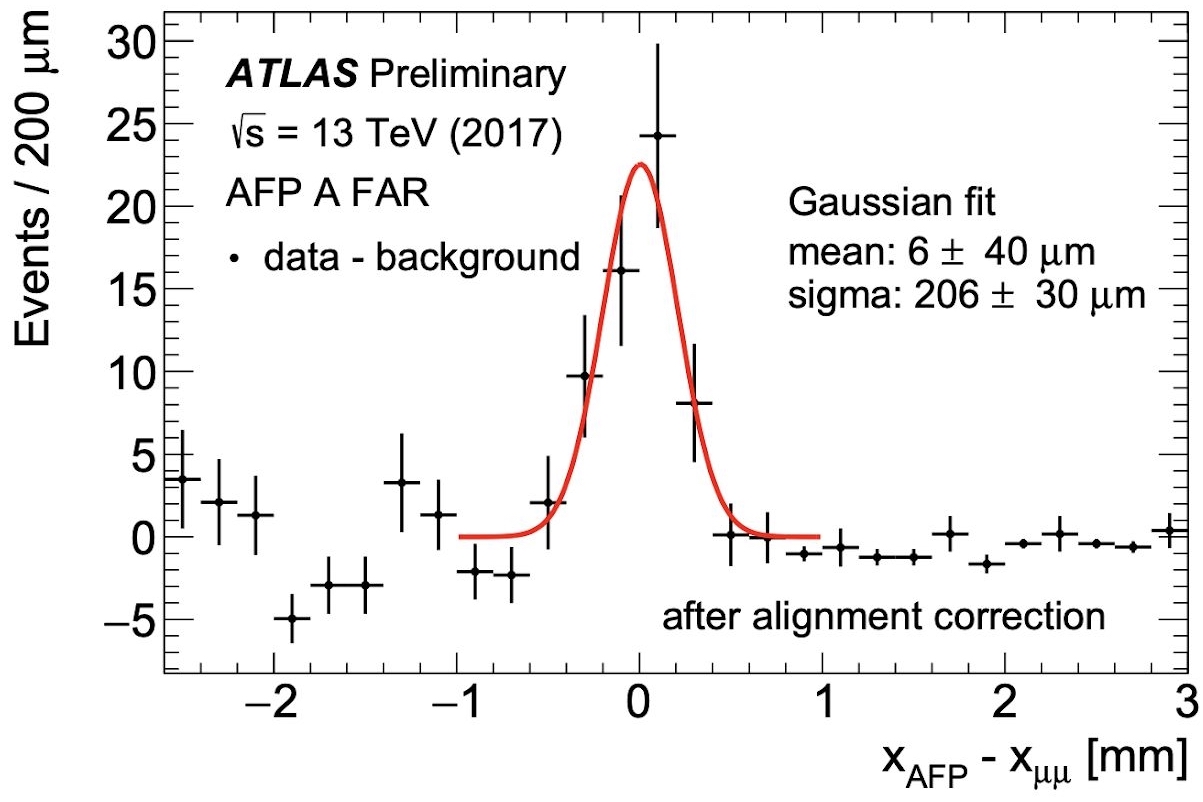}
\end{center}
\vspace*{-0.8cm}
\caption{Hit map recorded by the SiT detector (left), detector position before (center) and after global AFP alignment (right)~\cite{AFPpage}.
}
\label{fig:hitmap}
\vspace*{-0.7cm}
\end{figure}

The alignment of the SiT was performed with respect to the actual beam (later, it is planned with respect to the beam-pipe).
The 300\,$\mu$m mean of alignment over stations and data-taking is taken as systematic uncertainty
(Fig.~\ref{fig:hitmap}~\cite{AFPpage}).

%https://indico.cern.ch/event/868940/contributions/3813694/

\clearpage

\section{Trigger}
The AFP is part of the ATLAS data acquisition system.
The trigger rates are sent from the AFP detector at a nominal 20 sigma position from the beam.
Figure~\ref{fig:trigger}~\cite{AFPpage} shows the trigger rate as a function of the number of colliding bunches during LHC luminosity ramp-up
for 2015-2016 data-taking.

\section{ToF vertex reconstruction}
In the 2017 data, the expected and measured vertex resolution is about 5\,mm
(Fig.~\ref{fig:trigger}~\cite{AFPpage}).
% https://twiki.cern.ch/twiki/view/AtlasPublic/ForwardDetPublicResults
First results with the 2022 data have been obtained operating in Run-3 data taking.
% thesis Viktoria

\section{SiT track with ToF single train signal}
\begin{figure}[htb]
\vspace*{-0.3cm}
\begin{center}
\includegraphics[width=0.35\textwidth]
{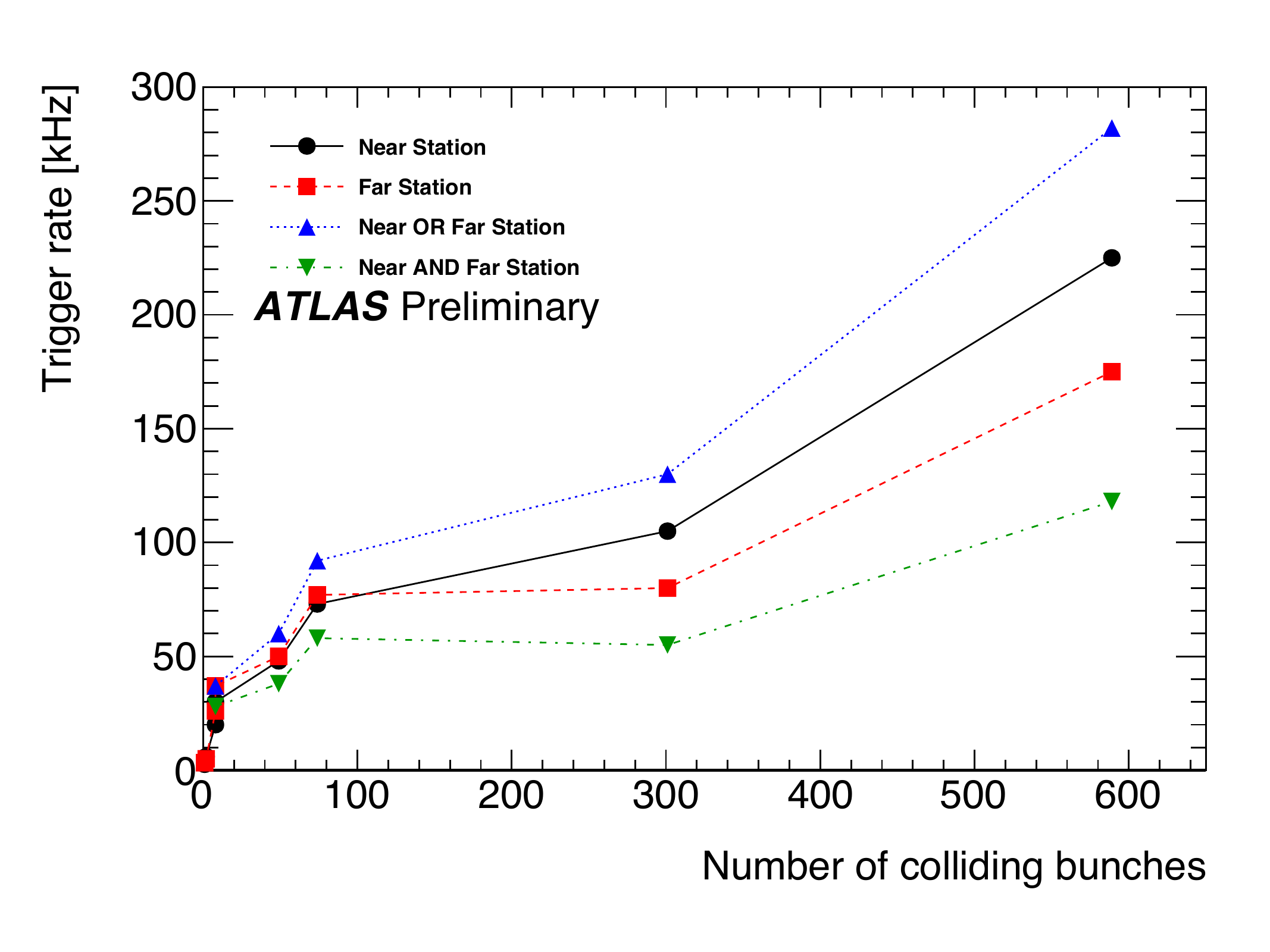}\hfill
\includegraphics[width=0.28\textwidth,angle=0]
{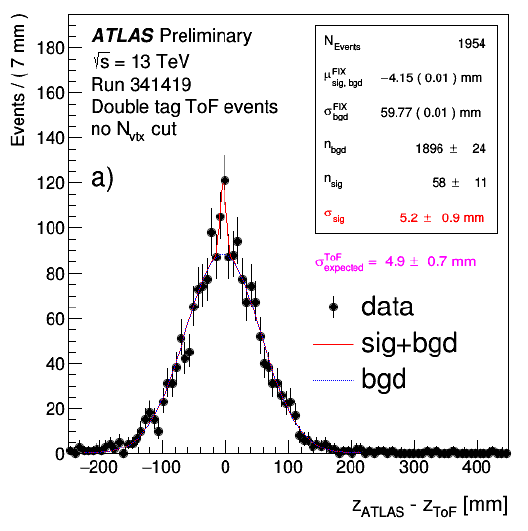}
\includegraphics[width=0.35\textwidth]
{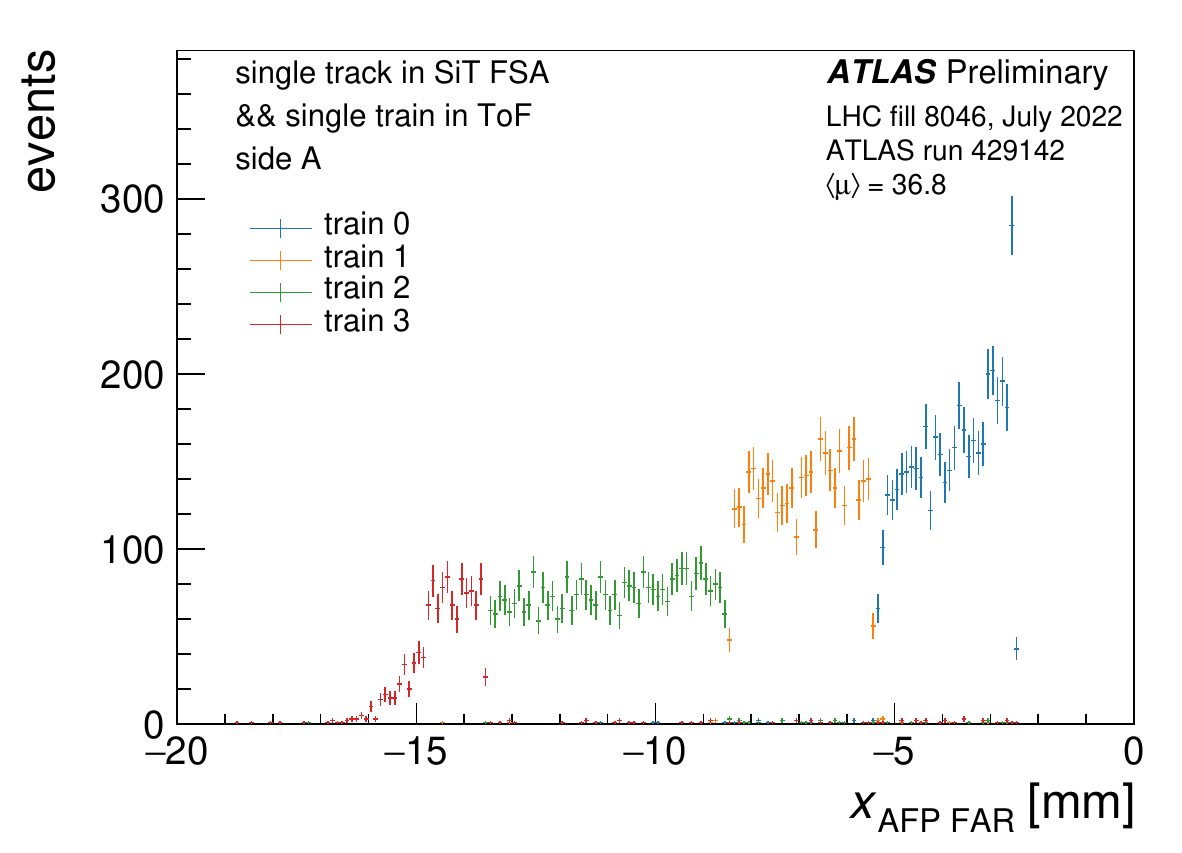}
\end{center}
\vspace*{-0.8cm}
\caption{
Trigger (left), ToF resolution (center) and SiT-ToF correlation (right)~\cite{AFPpage}.
}
\label{fig:trigger}
\vspace*{-0.7cm}
\end{figure}

The SiT-ToF correlation was determined by studying the $x$ position of tracks reconstructed in the SiT detector (FAR station) in events with a single-train signal in the ToF detector.
Different colors are used to visualize the SiT regions corresponding to individual trains
(Fig.~\ref{fig:trigger}~\cite{AFPpage}).
The machined $x$ widths of the ToF bars are 3/3/5/5.5\,mm for train 0/1/2/3.

% https://twiki.cern.ch/twiki/view/AtlasPublic/ForwardDetPublicResults

\section{SiT correlation with central ATLAS inner detector tracker/calorimeters}
\begin{figure}[htb]
\vspace*{-0.3cm}
\begin{center}
\includegraphics[width=0.49\textwidth]
{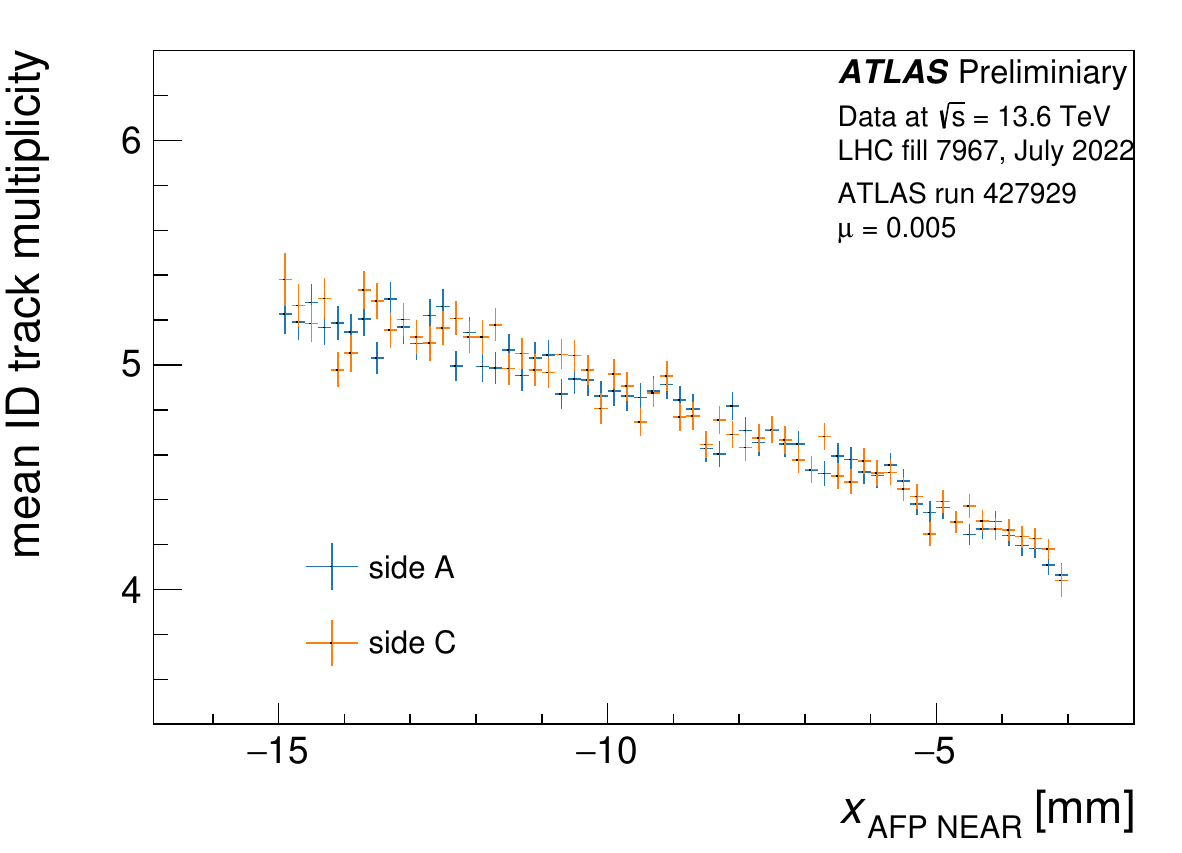}\hfill
\includegraphics[width=0.49\textwidth]
{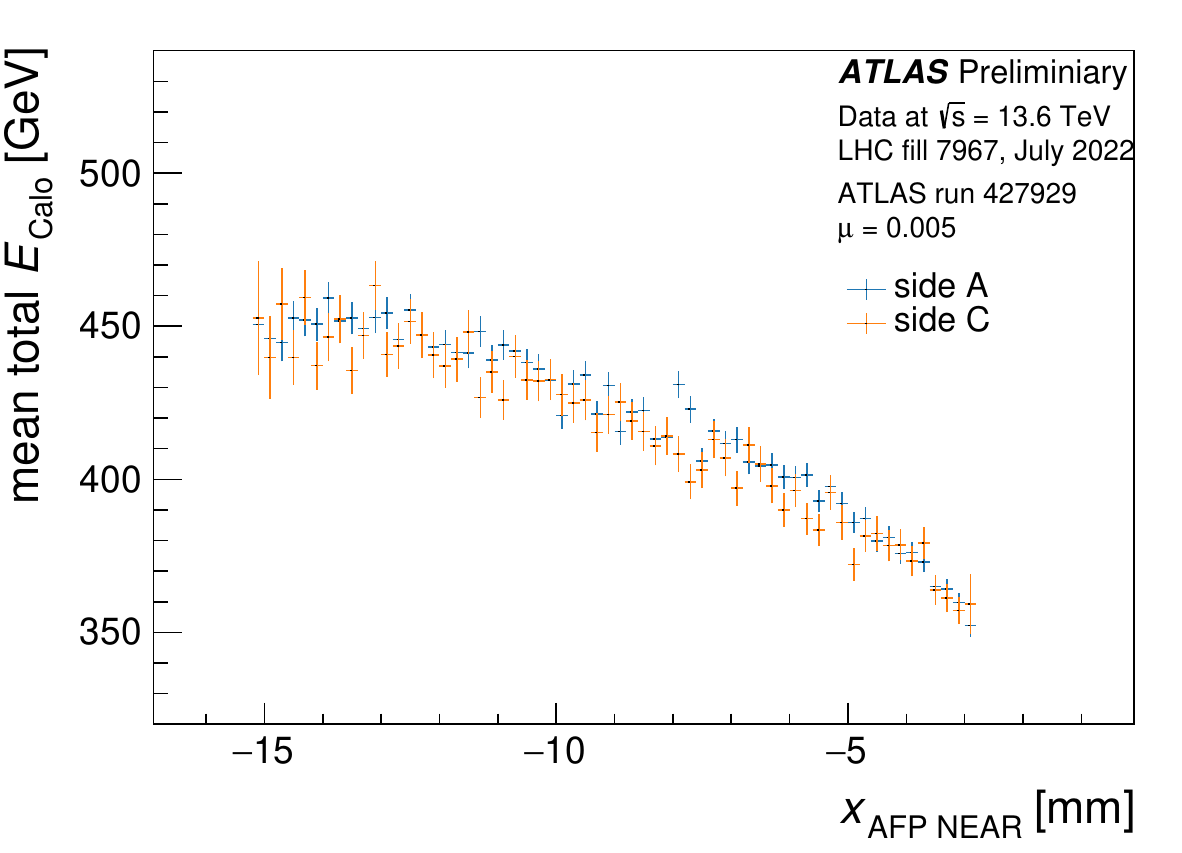}
\end{center}
\vspace*{-0.8cm}
\caption{
SiT correlation with ID tracker (left), SiT  correlation with calorimeters (right). The statistical uncertainty of the mean values for each bin are shown~\cite{AFPpage}.
}
\label{fig:sitidcalo}
\vspace*{-0.7cm}
\end{figure}

The correlations are shown between the $x$ position of reconstructed tracks in the AFP NEAR stations with a proton on side A or C and the charged track multiplicity in the ATLAS Inner Detector (ID), as well as the correlations with the total energy measured by the ATLAS calorimeters (Fig.~\ref{fig:sitidcalo}~\cite{AFPpage}). 
Exactly one reconstructed AFP track is required in each station and side, and a reconstructed primary vertex is required.
The ID track selection requires $pT> 500$\,MeV and $|{\eta}| < 2.5$.
Events with smaller $x$ AFP values are further away from the beam and are thus expected to originate from protons with higher energy loss.

% https://twiki.cern.ch/twiki/view/AtlasPublic/ForwardDetPublicResults

\section{Matching of proton energy loss with ATLAS central di-leptons/di-photons}
\begin{figure}[htb]
\vspace*{-0.3cm}
\begin{center}
\includegraphics[width=0.60\textwidth]
{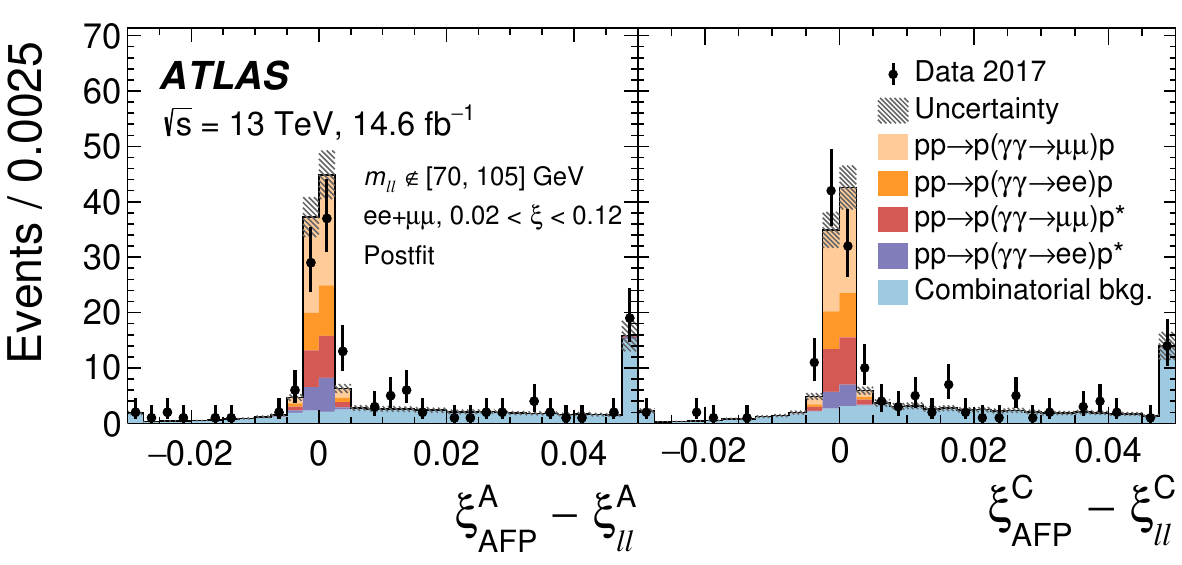}\hfill
\includegraphics[width=0.39\textwidth]
{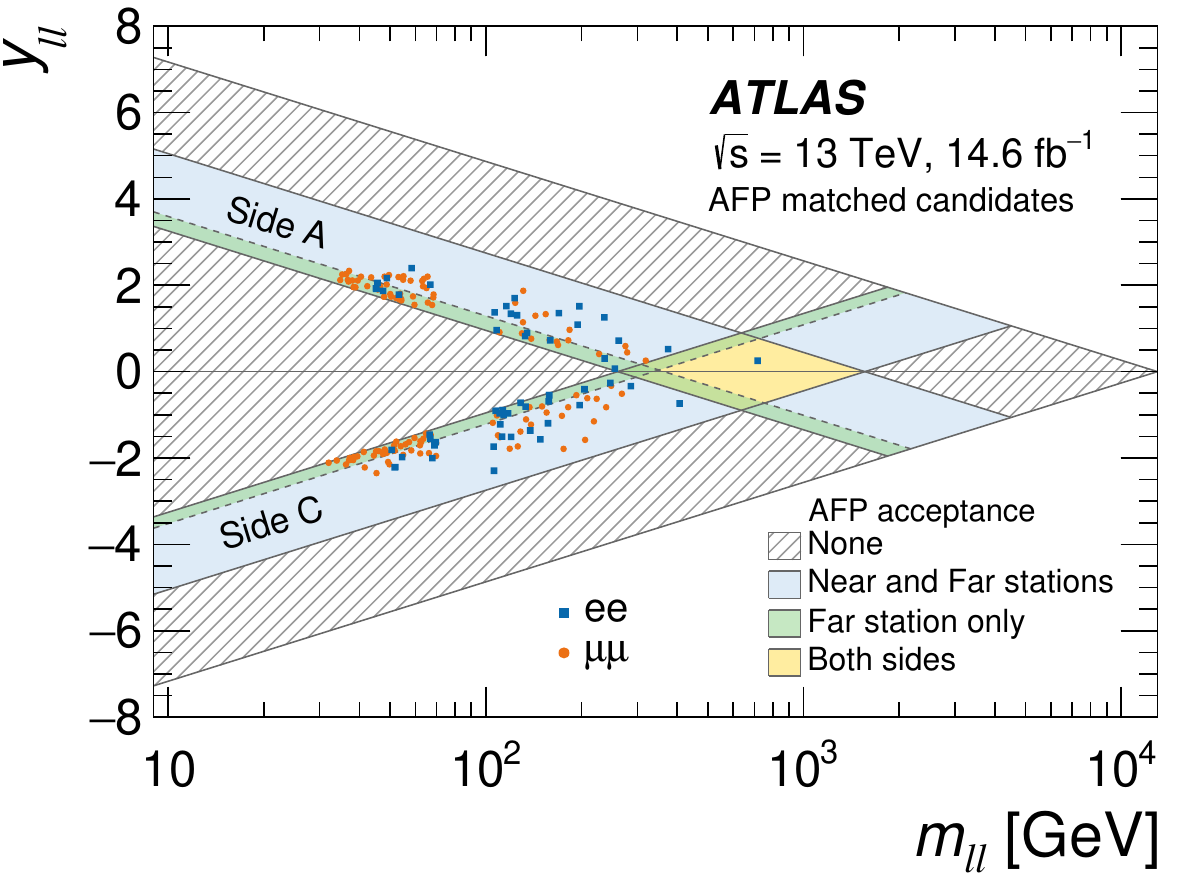}
\end{center}
\vspace*{-0.8cm}
\caption{
Di-lepton matching with AFP proton kinematics (left and center), di-lepton selected events (right)~\cite{Aad:2736159}.
%p* dissociated proton
}
\label{fig:matching}
\vspace*{-0.7cm}
\end{figure}

The photon-induced di-lepton production with forward proton tag at 13\,TeV
was studied in the AFP acceptance range $0.02 < \xi < 0.12$, where $\xi$ is the relative proton energy loss~\cite{Aad:2736159}. The
signal and combinatorial background processes are shown in Fig.~\ref{fig:matching}~\cite{Aad:2736159}.
Di-lepton events are studied in the rapidity $y_{\ell\ell}$ versus $m_{\ell\ell}$ plane
using 14.6\,fb$^{-1}$~\cite{Aad:2736159}.
Event are selected with the kinematic matching 
$|\xi_{\rm AFP}-\xi_{\ell\ell}| < 0.005$ on at least one side (Fig.~\ref{fig:matching}~\cite{Aad:2736159}).
Shaded (hatched) areas denote the acceptance (no acceptance) for the AFP stations.
Areas neither shaded nor hatched correspond to $\xi$ outside [0,1].

For the light-by-light scattering mediated Axion-Like-Particle (ALP) production, the matching of a photon pair and the proton kinematics is required.
Figure~\ref{fig:ALP}~\cite{Aad:2856777} shows 441 single matching events in 2017 data. There are no double matching events. The matching requirement is $|\xi_{\rm AFP}-\xi_{\gamma\gamma}| < 0.004 + 0.1\xi_{\gamma\gamma}$.
%ATLAS-CONF-2023-002, JHEP

\clearpage

\section{Key physics results}
For the $\gamma\gamma\rightarrow \ell\ell$ analysis, 
57 (123) candidates $\rm e^+e^- +p (\mu\mu +p)$ events are selected~\cite{Aad:2736159}.
The background-only hypothesis is rejected with a significance $>5\sigma$ in each channel.
Cross-section measurements in the fiducial detector acceptance $\xi \in [0.035,0.08]$ yield: \\
$\rm \sigma (ee+p)=11.0\pm2.6 (stat) \pm1.2 (syst) \pm0.3 (lumi)$ fb, and \\
$\rm \sigma (\mu\mu+p)=7.2\pm1.6 (stat)\pm0.9 (syst)\pm0.2 (lumi)$ fb.\\
A comparison with proton soft survival (no additional soft re-scattering) models gives:\\
$10.0\pm0.8$~fb (ee+p) and $9.4\pm0.7$~fb ($\mu\mu$+p)~\cite{Aad:2736159}.

\begin{figure}[htb]
\vspace*{-0.3cm}
\begin{center}
\includegraphics[width=0.29\textwidth]
{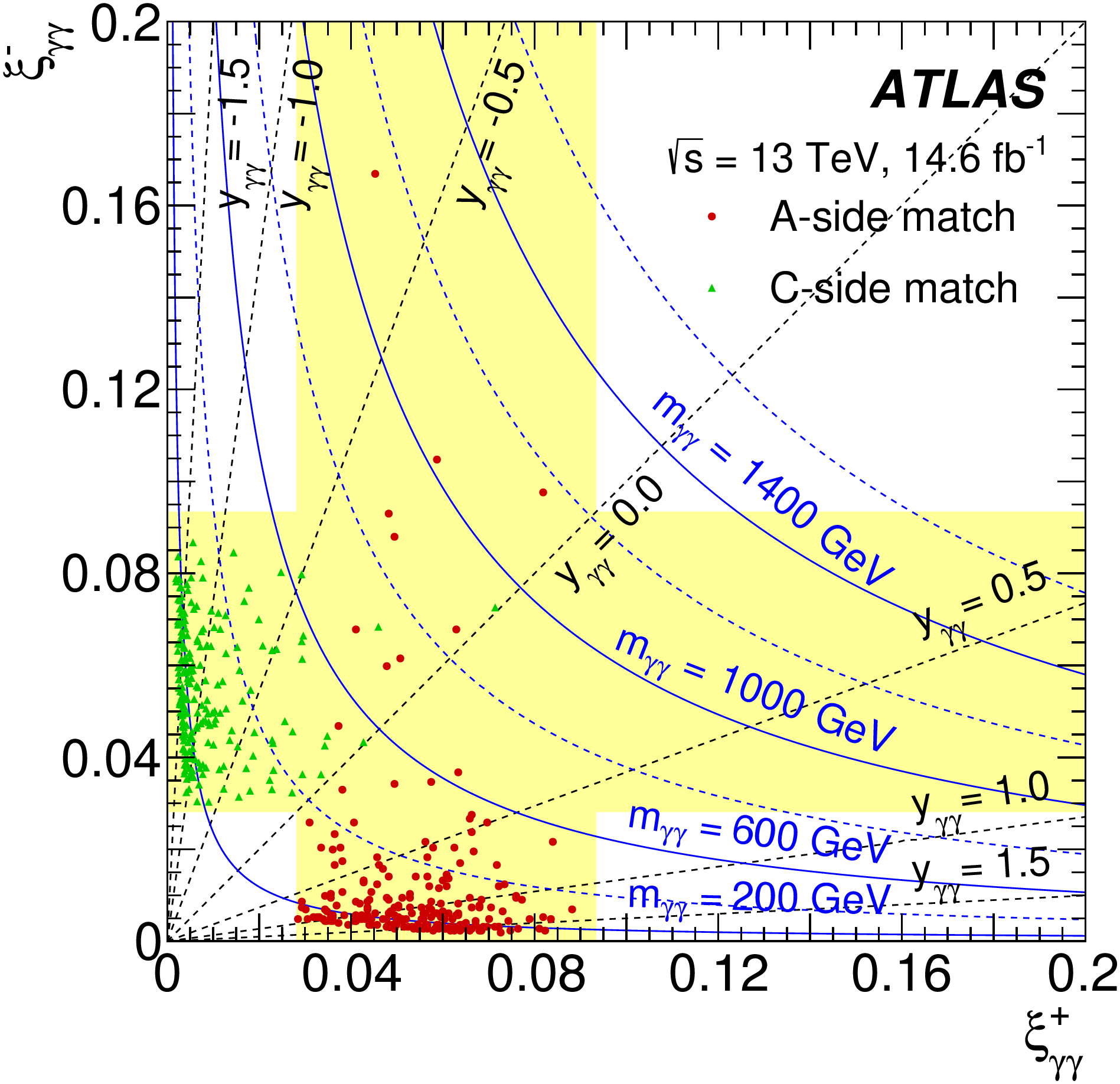}\hfill
\includegraphics[width=0.35\textwidth]
{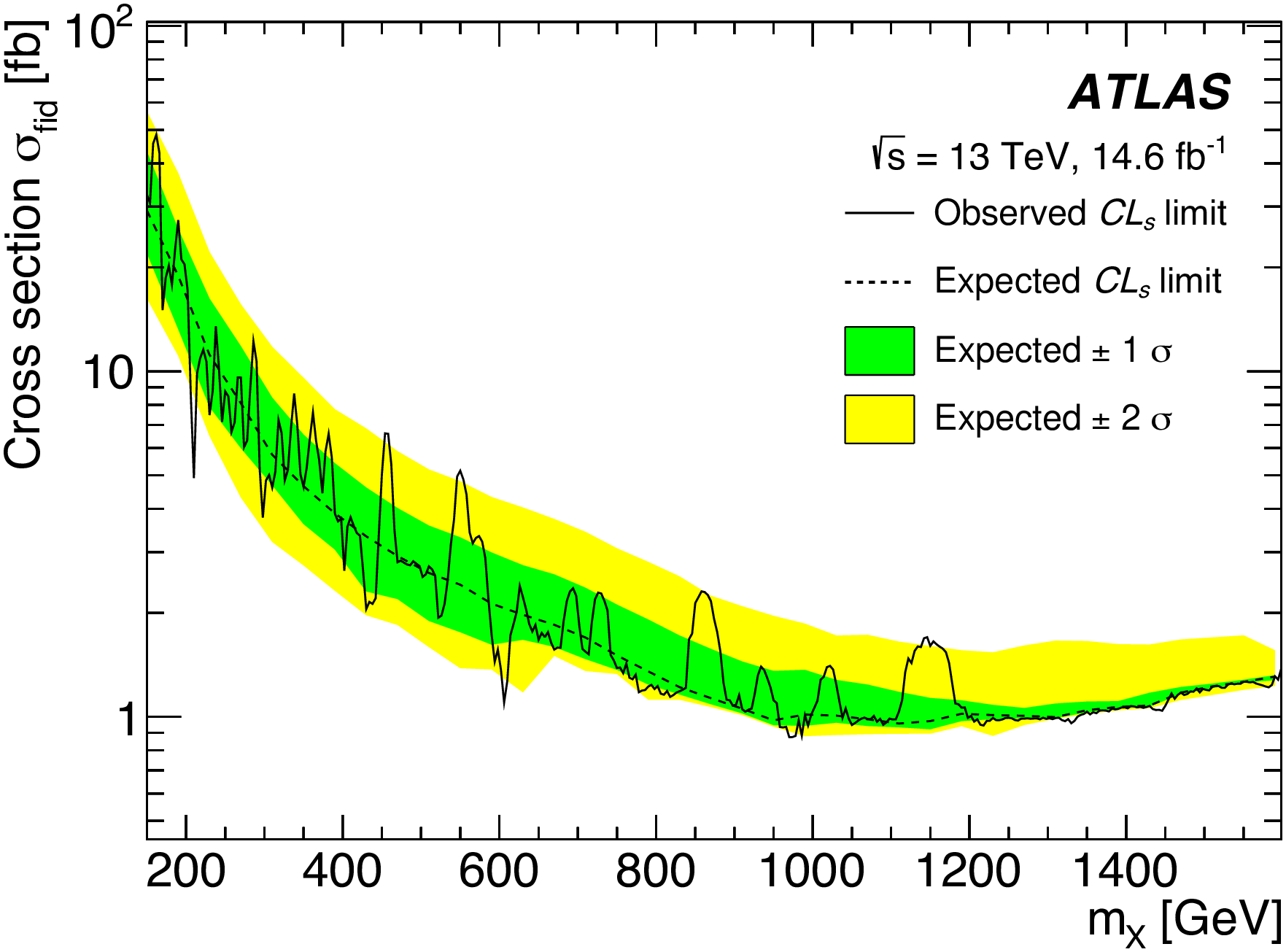}
\includegraphics[width=0.35\textwidth]
{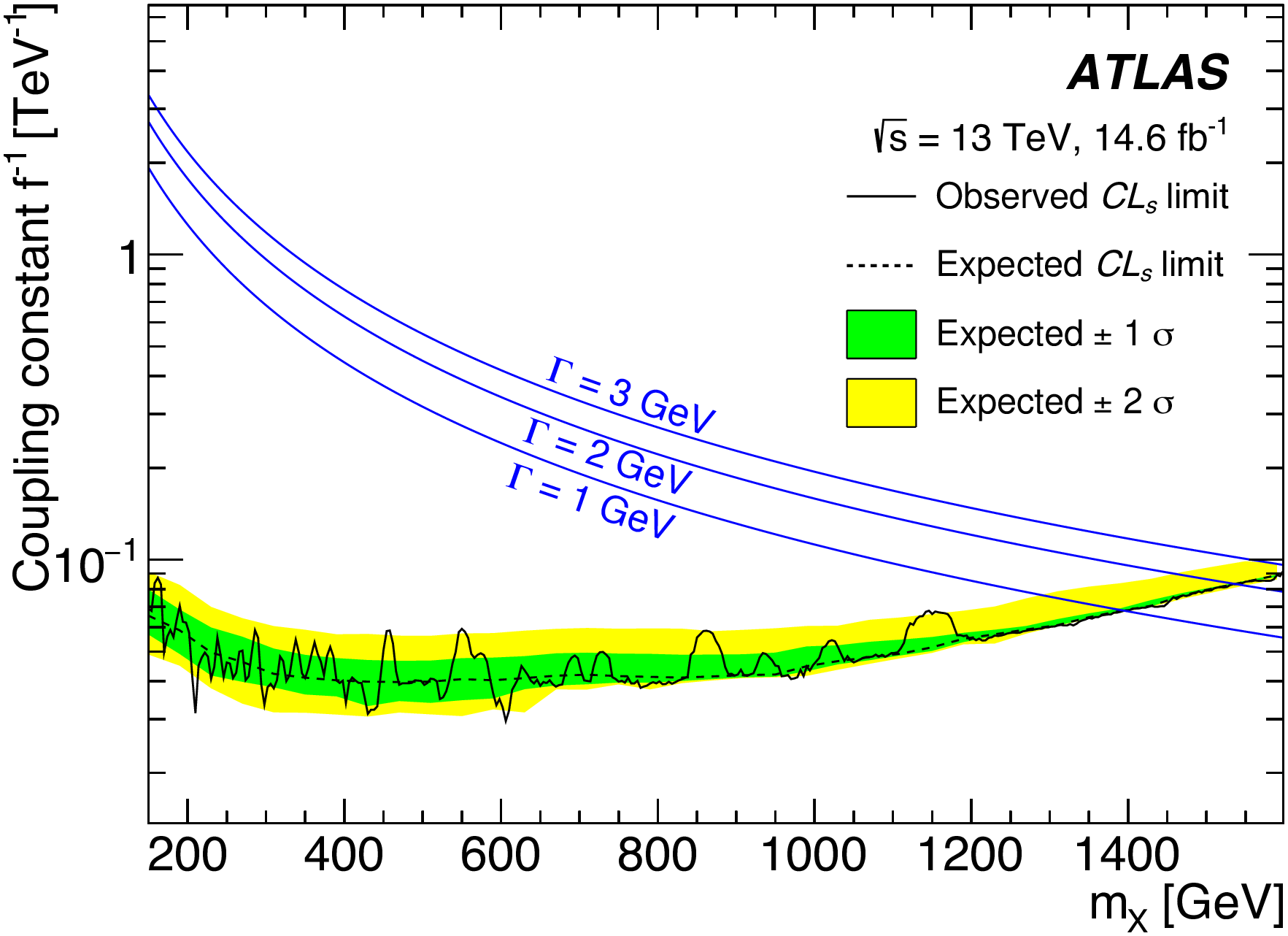}
\end{center}
\vspace*{-0.8cm}
\caption{
441 di-photon events with AFP tag (left), 
limit on ALP production cross-section (center) and ALP coupling (right)~\cite{Aad:2856777}.
}
\label{fig:ALP}
\vspace*{-0.7cm}
\end{figure}

The Axion-Like-Particle (ALP) search in the reaction 
$\gamma\gamma\rightarrow \gamma\gamma$ uses an AFP proton tag to reduce the background, and it
leads to limits on the ALP production cross-section and ALP coupling (Fig.~\ref{fig:ALP}~\cite{Aad:2856777}).
There is a further rich analysis programme using the AFP detector~\cite{AFPpage}.

\section{Outlook for Run-3}
The LHC Run-3 proton-proton interactions have a large potential for the AFP detector operation.
There are several LHC operation modes:\\
$\bullet$ High-$\mu$ runs for regular data taking  (AFP integrated luminosity expected to match ATLAS reference), purpose: high $pT$ exclusive processes and Beyond the Standard Model searches.\\
$\bullet$ Low-$\mu$ runs (various pile-up conditions, $0.005<\mu<1$) during LHC ramp-ups, purpose: soft diffraction, low $pT$ hard diffraction.\\
$\bullet$ ``LHCf run $\beta = 19$\,m", purpose: diffractive studies, connection to cosmic ray physics.\\
$\bullet$ Medium-$\mu$ runs ($\mu \approx 2$) with 1\,fb$^{-1}$ of data planned to be collected, purpose: excellent sample to study medium/high $pT$ hard diffractive processes. The process $\rm pp~\rightarrow~PbPb$ will be used as reference run for diffractive studies at lower energy.

\vspace{-3mm}
\section{Conclusions}
\vspace{-1mm}
The ATLAS Forward Proton (AFP) detector has operated successful at LHC Run-2 and Run-3 for both silicon Tracker (SiT) and
Time-of-Flight (ToF).
Single dissociative (soft QCD) and re-scattering were probed in photon-induced di-lepton production with a forward proton tag.
Light-by-light scattering were used in Beyond the Standard Model searches using a forward proton tag.

Future precision increases with the analysis of Run-3 data is anticipated using larger data sets (2022-2025) and using the ToF data for improved vertex reconstruction and further background rejection.

\vspace{-3mm}
\section*{Acknowledgments}
\vspace{-1mm}
The project is supported by the 
Ministry of Education, Youth and Sports of the Czech Republic under project number LM\,2015058 and LTT\,17018.

Copyright 2023 CERN for the benefit of the ATLAS Collaboration. CC-BY-4.0 license.

\vspace{-3mm}
\bibliographystyle{unsrt}

\bibliography{DIS2023}

\begin{thebibliography}{1}

\bibitem{AFPpage}
{{ATLAS Collaboration}}.
\newblock {Public Forward Detector Plots for Collision Data}.
\newblock
  {https://twiki.cern.ch/twiki/bin/view/AtlasPublic/ForwardDetPublicResults}.

\bibitem{LumiRun3}
{{ATLAS Collaboration}}.
\newblock {Public ATLAS Luminosity Results for Run-3 of the LHC}.
\newblock
  {https://twiki.cern.ch/twiki/bin/view/AtlasPublic/LuminosityPublicResultsRun3}.

\bibitem{Aad:2736159}
{{ATLAS Collaboration}}.
\newblock {Observation and measurement of forward proton scattering in
  association with lepton pairs produced via the photon fusion mechanism at
  ATLAS}.
\newblock {\em Phys. Rev. Lett.}, 125(26):261801, 2020.

\bibitem{Aad:2856777}
{{ATLAS Collaboration}}.
\newblock {Search for an axion-like particle with forward proton scattering in
  association with photon pairs at ATLAS}.
\newblock 2023.
\newblock arXiv:2304.1095335, accepted JHEP.

\end{thebibliography}
\vspace{-3mm}
\end{document}